\documentstyle[preprint,aps,psfig]{revtex}
\begin{document}
\draft
\preprint{
\vbox{
\hbox{TUM/T39-96-12}  
\hbox{Februar 1996}
}}
\title{MC simulation of vector boson 
       production in polarized 
       proton collisions}
\author{A. Saalfeld$^{a,b}$ and A. Sch\"{a}fer$^{b}$}
\address{$^a$   Physik Department,
                Technische Universit\"{a}t 
                M\"unchen,
                D-85747 Garching, Germany}
\address{$^b$   Institut f\"{u}r 
                theoretische Physik,
                Universit\"{a}t Frankfurt,
                D-60054 Frankfurt am Main, 
                Germany}
\maketitle
\begin{abstract}
We present a Monte Carlo (MC) study of $W^{\pm}$ 
and $Z_{0}$ production of longitudinally 
polarized proton collisions at a center 
of mass (CMS) energy of $500\,{\rm GeV}$. 
All results were obtained with the 
SPHINX MC code. We consider two 
different types of polarized parton 
distributions with large and
small gluon polarization.
Various spin asymmetries are 
found to be sensitive to the shape 
of the polarized gluon distribution between 
$0.1\stackrel{<}{\sim}x\stackrel{<}{\sim}0.5$.
The asymmetries are approximately
constant with respect to the transverse momentum 
carried by the $W^{\pm}$ and $Z_{0}$, 
{\em except} if the first moment of the
gluon polarization is large. 
In this case some $W^{\pm}$ asymmetries 
show a significant variation with transverse 
momentum. 
\end{abstract}
\pacs{PACS numbers: 13.60Hb, 13.88.+e, 14.20.Dh
\\  
Submitted to {\em Phys.Rev.D}}
%
\section{Introduction}
In the past few years polarized deep 
inelastic scattering became a very attractive 
topic in high-energy physics.
Experimentally, an important advance has been 
the measurement of the structure function $g^p_1$
of the proton with high statistics 
\cite{ashman88a,adams94a}.   
On the theoretical side much effort has been done 
to decompose $g^p_1$ into parton 
distributions, but the situation is still unclear
(see, e.g., \cite{anselmino95a} for a recent review). 
This is especially true for the polarization of 
the sea quarks, the polarization of the gluons,
the shape of the parton distributions at 
intermediate Bj{\o}rken $x$, and the small-$x$ 
behavior of the parton distributions. 
\\
There are in principle means to calculate all 
these quantities either from QCD sum rules or from 
lattice QCD. It will be exciting to compare 
future calculations and experiments.
\\
Meanwhile the parton distributions are 
constructed from models of the nucleon, 
e.g., from Carlitz-Kaur type models 
\cite{ehrnsperger95b} and many others.
An even more heuristic way to obtain the
parton distributions is simply to assume    
a generic shape of the polarized distributions
and fit the strength and size parameters
to the $g_1$ data. This
procedure is of course in no way unique.
Many of these distributions are motivated from
properties of the unpolarized distributions at 
larger $x$. Some fits invoke an SU(3) symmetric
polarized sea in order to fit the first moments
of the parton distributions in terms of 
$\beta$-decay and hyperon decay matrix elements,
see, e.g., \cite{gehrmann95a}.
\\
The naive way of comparing model parton 
distributions with parton distributions 
defined through expectation 
values of field theoretic
operators has been much criticized 
\cite{carlitz88a,mankiewicz90a}. 
It has been argued that a proper comparison 
can only be made after the axial anomaly 
has been taken into account which shifts 
the first moments of the naive parton 
distributions against the first moments
of the field theoretic parton distributions by
\begin{equation}
\label{anomalycontrib}
\Delta q_f=
  \tilde {\Delta} q_f -
  \frac{\alpha_s}{2{\pi}}\Delta{g}, 
\end{equation}
where $\Delta q_f$ denotes the field
theoretic distribution of a quark 
with flavor f and $\tilde {\Delta} q_f$
denotes the corresponding quantity within 
a quark model. 
$\Delta g$ is the first moment
of the polarized gluon 
distribution. Equation (\ref{anomalycontrib})
is valid for any light quark flavor f.
This equation is special for 
longitudinal polarization.  
Analyzing the data within a quark model, 
one has to fix the four unknown quantities 
$\tilde {\Delta} u$, $\tilde {\Delta} d$,
$\tilde {\Delta} s$, and $\Delta g$, neglecting
the contribution of heavy quarks.
But even when assuming an SU(3) symmetric
sea quark distribution, the $g_1$, 
$\beta$-decay, and hyperon decay experiments
only fix three out of the four unknown 
first moments. The shapes of the distributions
are of course not known at all from these 
considerations.
\\
Therefore other experiments have been proposed
to complement the measurement of $g_1$ of the
proton. The most important ones being 
semi-inclusive deep inelastic scattering 
\cite{gullens93a} and polarized proton scattering 
\cite{rhic92a,rhic93a}. 
In the present work we concentrate on proton 
scattering. The production of large 
transverse momentum photons in polarized proton
collisions might be a suitable observable, but
there is a large background from pion decays 
in the final state, as was investigated 
in detail in ref. \cite{gullens94b}. This
is not so for $\rm{W}^\pm$ and 
${\rm Z}_0$ production 
at RHIC where the background from other events
will be very small \cite{rhic92a}. The limiting
factor in this experiment will be the number of  
$\rm{W}^\pm$ and ${\rm Z}_0$ that can be produced.
There are at least three different poorly known
properties of the partons, which must be 
entangled in future experiments, namely,  the 
amount of flavor breaking in the sea quark 
distribution, the size of the gluon distribution, 
and the shape of all these distributions.
In this paper we only consider two  
different parametrizations.
The first parametrization stems from 
a Carlitz-Kaur type constituent 
quark model with a polarized gluon, 
radiatively generated sea quarks 
starting from ${\rm Q}^2=10\,{\rm GeV}^2$,
and a slightly broken ${\rm SU(2)}
\times {\rm SU(3)}$ symmetry 
\cite{ehrnsperger95b}. This model does
not use the anomaly equation 
(\ref{anomalycontrib}),
but rather identifies $\Delta q_f$ with 
$\tilde {\Delta} q_f$. 
This can be justified a posteriori because
in this model the gluon polarization 
$\Delta g=0.262 $ turns out to be small.
The interesting feature of these distributions
is that they satisfy the Bj{\o}rken sum rule 
{\em and} the Ellis-Jaffe sum rule.
\\  
The other parton distributions have been 
taken from ref. \cite{gehrmann95a}. These 
distributions also have SU(3) symmetric
sea quarks, but the sea quarks are 
somewhat stronger polarized 
due to the lower  starting 
point  ${\rm Q}^2=4\,{\rm GeV}^2$ 
of the evolution. The analysis in 
ref. \cite{gehrmann95a} proceeds 
via the anomaly equation 
(\ref{anomalycontrib})
and a larger ratio $F/D=0.590$. Therefore,
a much larger gluon polarization
of $\Delta g = 1.971$ at 
${\rm Q}^2=4\,{\rm GeV}^2$
is obtained. The basic question we want 
to discuss is: Can one distinguish between these
parametrizations in a $W^\pm/Z_0$ experiment, e.g., 
at RHIC?
\\  
The measurement of the flavor breaking 
in the sea quark distribution by studying 
$\rm{W}^\pm$ and $Z_0$ production is not addressed 
in this paper. A discussion of this interesting 
subject can be found in ref. \cite{bourrely94a}.
\\
The paper is organized as follows. In section II
we give a very brief review of the SPHINX MC code 
\cite{gullens94b,gullens94a}, in which the 
electroweak matrix elements have now been included.
In section III we present the parton distributions
that are used in our MC simulation. Finally,
in section IV we present and  discuss
the results.
%
\section{The Monte Carlo program SPHINX} 
\label{sphinx}
The application of perturbative QCD to Drell-Yan
like reactions is complicated through the possible
emission of soft and collinear radiation of 
partons in the initial and final state. The 
importance of these bremsstrahlung effects is
seen most clearly from the experimental transverse
momentum spectra of ${\rm W}^\pm$ and 
${\rm Z}_0$ production cross sections. These  
spectra cannot be explained through an intrinsic
transverse momentum of the initial partons. We
must rather think of the transverse momentum as 
being generated by bremsstrahlung effects.
It is not possible to calculate the K-factor
exactly to arbitrary order $\alpha_s$, so 
at some stage one must rely on approximations.
\\
The MC generator SPHINX \cite{gullens94b},
which is an extension of the well-known PYTHIA 
5.6 MC code \cite{sjostrand86a}, treats parton 
radiation in leading logarithmic approximation 
(LLA). This means that it resums the planar 
bremsstrahlung graphs in an axial gauge. 
These graphs are exactly the 
universal bremsstrahlung contributions 
that do not depend on the particular 
process, e.g., 
$u {\overline d} \rightarrow {\rm W}^+$ or
$u {\overline u} \rightarrow \gamma/{\rm Z}_0$.
In SPHINX and PYTHIA the bremsstrahlung cascade
is resummed in backwards direction, starting from 
the most virtual partons immediately before the
reaction takes place. The longitudinal momenta
of the mother partons are reconstructed
using first order polarized GLAP 
equations and evolved parton distributions.
All information from the GLAP equation 
necessary to reconstruct the longitudinal
momenta of the mother partons is contained 
in the inverse Sudakov-like formfactor
\begin{eqnarray}
\label{inversesudakov}
{\rm S}_{b,h_b}(x,t_<,t_>)=exp\left(
-\int_{t_<}^{t_>} dt' 
\frac{\alpha_s(t')}{2\pi}\sum_{a,h_a}
\int \frac{dx'}{x'} 
\frac{x'f_{a, h_a}(x',t')}{xf_{b,h_b}(x,t')}   
P_{a,h_a\rightarrow b,{h_b},c,{h_c}}(\frac{x}{x'})
\right),
\end{eqnarray}
giving the probability that the parton remains at
$x$ from $t_>=ln(Q_{>}^2/\Lambda^2)$ to   
$t_<=ln(Q_{<}^2/\Lambda^2)$, where $Q_{<}^2$ and 
$Q_{>}^2$ denote the virtualities  of the
daughter and mother partons respectively.
The indices $a$ and $b$ denote the flavors of
the partons and $h_a$ and $h_b$ their 
helicities. The polarized splitting functions
$P_{a,h_a\rightarrow b,h_b,c,h_c}(z)$ can be found 
in ref.\cite{altarelli77a}. As it stands, the 
integrals diverge for the emission of soft radiation
due to the poles in the splitting functions. 
To cure this, the soft region in the integral 
is cut out and treated separately. 
This causes no problem \cite{sjostrand86a}.
Note, that SPHINX cannot evolve parton 
distributions but rather needs evolved 
partons as can be seen from equation 
(\ref{inversesudakov}). 
For the leading-order GLAP evolution we used 
the code developed by Gehrmann
and Stirling \cite{gehrmann95a}.
\\
The full four-momenta of the mother partons
can be reconstructed recursively: Consider   
two partons with four-momenta $p_1, p_2$,
and virtualities $Q_1^2=-p_1^2$ and 
$Q_2^2=-p_2^2$ in their common CMS frame. 
Let $Q_1^2>Q_2^2$. Then the momentum of
the mother parton $3$ of parton $1$ 
is reconstructed while parton $2$
is not changed. Finally, parton $3$ and parton $2$
are boosted to their common CMS frame and the 
procedure starts again until both partons have
a virtuality smaller than $1$\,GeV.   
\\
As is well known, the variables $x$ 
used in the splitting functions and
in the parton distributions have no unique 
kinematical interpretation in LLA, see, e.g., 
\cite{dokshitser80a}. For technical reasons
in SPHINX the same prescription as in PYTHIA 
is used, namely, the ${\hat s}$-approach where
${\hat s} = x_1 x_2 s$ is the momentary CMS energy
of to partons at any stage of the evolution
and $s$ denotes the overall CMS energy 
of the reaction.
\\
The last ingredients to  be specified are 
the hard scattering cross sections. At low
transverse momentum a weak boson is produced 
through a simple Breit-Wigner resonance. In
SPHINX the treatment of the s-dependent
width of the resonance is the same as in 
PYTHIA 5.6. We simply use that the 
polarized hard scattering cross sections 
are related to the unpolarized ones by
\begin{equation}
\label{wresonance}
{\hat\sigma}_{W^\pm}({\hat s},h_1,h_2)=
\frac{1}{2} 
\delta_{h_{1}h_2,-1} \; 
{\hat\sigma}_{W^{\pm}}^{unpol}({\hat s}), 
\end{equation}
where $h_1$, and $h_2$ are the helicities
of the quark and antiquark respectively.  
A similar relation is true for $Z_0$ production
\begin{equation}
\label{zresonance}
{\hat\sigma}_{Z_0}({\hat s},h_1,h_2)=
\delta_{h_1 h_2,-1} \;
\left\{\delta_{h_1,+1}\frac{C_L^2}{C_L^2+C_R^2}+
\delta_{h_1,-1}\frac{C_R^2}{C_L^2+C_R^2}\right\}
{\hat\sigma}_{Z_0}^{unpol}({\hat s}) 
\end{equation}
with $C_L=g_v+g_a$ and $C_R=g_v-g_a$ expressed
through the axial and vector coupling constants
$g_a$, and $g_v$ respectively.
If the transverse momenta of the vector bosons
become large the collinear approximation in  
the treatment of initial state radiation
becomes bad and one should use higher order 
matrix elements in combination with 
the radiation algorithm. 
For this purpose we implemented the 
next-to-leading
order QCD and QED polarized cross sections 
in SPHINX. The QCD cross sections can be 
found in ref. \cite{hidaka81a} but 
we also list them  here: The annihilation 
process $q(h_1)+{\overline q}(h_2)
\rightarrow g+Vectorboson$
has to lowest order in $\alpha_s$ 
the scattering cross section     
\begin{eqnarray}
\label{annihilation}
\frac{d\hat{\sigma}_{q(h_{1}),
\overline{q}(h_{2})}}{d\hat{t}}=
\frac{2}{9}\alpha_{s}(b-h_1a)^2
(1-h_1h_2)\frac{1}{\hat{s}^2}
\left(\frac{\hat{u}}{\hat{t}}
+\frac{\hat{t}}{\hat{u}}
+\frac{2M_{V}^2\hat{s}}
{\hat{t}\hat{u}}\right).
\end{eqnarray}
The QCD Compton process
$q(h)+g(\lambda)\rightarrow 
q + Vectorboson$
has in leading order $\alpha_s$ 
the cross section
\begin{eqnarray}
\label{compton}
\frac{d\hat{\sigma}_{q(h),
g(\lambda)}}{d\hat{t}}=
-\frac{1}{12}\alpha_{s}(b-ha)^2 
\frac{1}{{\hat s}^2}
\left\{ (1+\lambda h) 
\left(\frac{\hat{s}}{\hat{t}} 
+\frac{\hat{t}}{\hat{s}}
+\frac{2M_{V}^2\hat{u}}{\hat{t}\hat{s}}
\right)-2\lambda h \frac{\left(M_{V}^2
-\hat{t}\right)^2}{\hat{t}\hat{s}}\right\}.
\end{eqnarray}
Here $h_1,h_2,h(=\pm 1)$, and $\lambda(=\pm 1)$
denote the incoming parton helicities. 
For $Z_0$ production one has to substitute 
\begin{eqnarray}
a_u&=&-a_d=\frac{g}{4cos\theta_W},
\nonumber\\
b_u&=&\frac{g}{4cos\theta_W}
\left(-1+\frac{8}{3}sin^2\theta_W\right),
\nonumber\\
b_d&=&\frac{g}{4cos\theta_W}
\left(1-\frac{4}{3}sin^2\theta_W\right),
\end{eqnarray}
for $a$ and $b$, where $g=e/sin \theta_{W}$.
In the case of $W^\pm$ production one has 
to substitute 
\begin{eqnarray}
a_{ud} &=& b_{ud} =-\frac {g} {2\sqrt{2}} 
cos\theta_{C},
\nonumber\\
a_{us} &=& b_{us} =-\frac {g} {2\sqrt{2}} 
sin\theta_{C},
\end{eqnarray}
for $a$ and $b$, where $\theta_{C}$ is the 
Cabbibo angle.
%
\section{The Parton distributions}
\begin{figure}
\centering{\psfig{figure=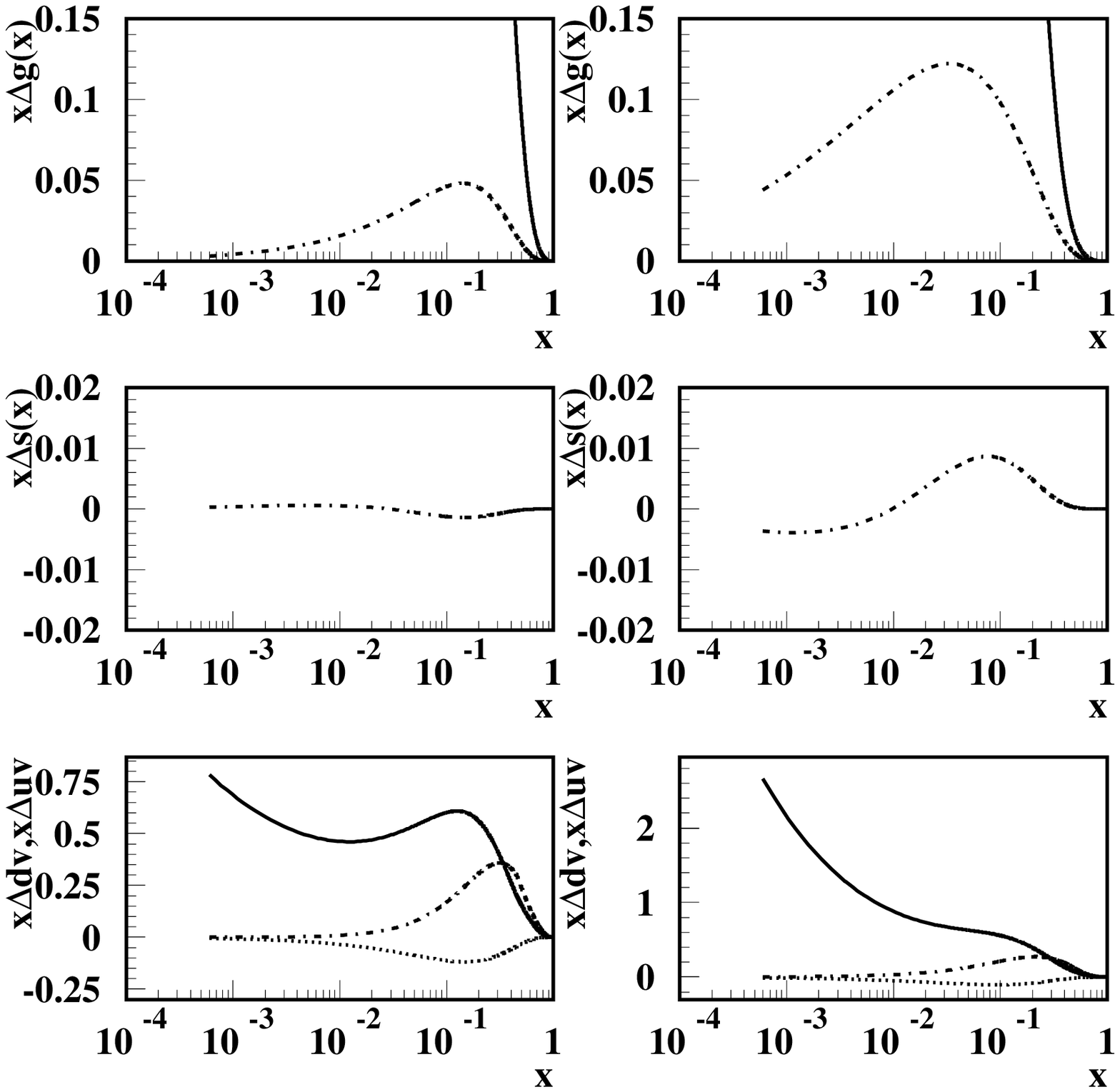,height=20cm}}
\caption{MC reconstruction of the 
CKT parton distribution 
with SPHINX at
$Q^2=4\,{\rm GeV}^2$ (left), and
$Q^2=8100\,{\rm GeV}^2$ (right). 
For comparison, also
the unpolarized quantities 
are shown (solid lines).}
\label{bruno}
\end{figure}

\begin{figure}
\centering{\psfig{figure=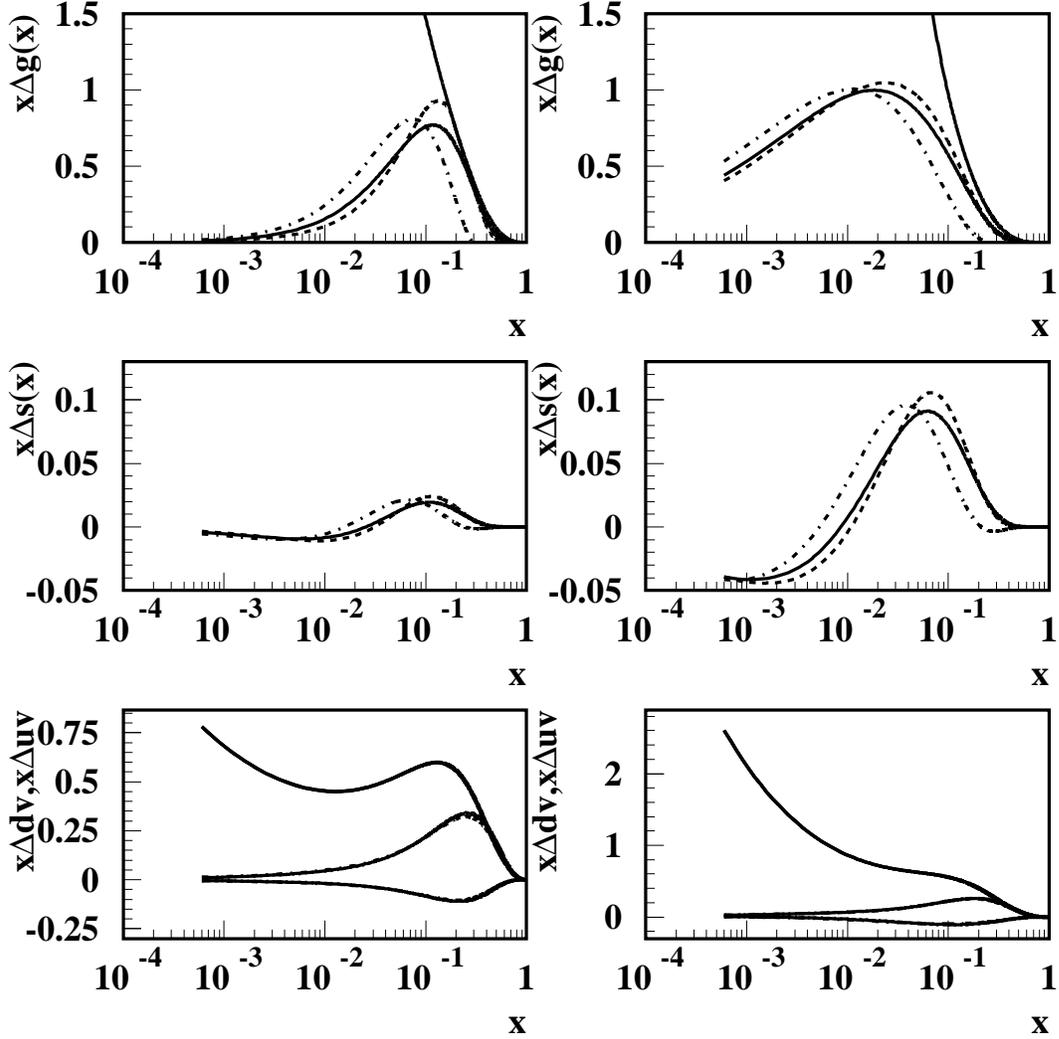,height=20cm}}
\caption{MC reconstruction of the parton 
distributions SETA, SETB, and SETC 
with SPHINX at
$Q^2=4\,{\rm GeV}^2$ (left), and
$Q^2=8100\,{\rm GeV}^2$ (right). 
Solid lines show the SETA distributions,
dashed lines SETB, and dash-dotted lines 
SETC. For comparison, also
the unpolarized quantities are shown 
(solid line).}
\label{geh} 
\end{figure}

In this section we describe the main physics
properties of the parton distributions 
that we used in the simulation. For details
see refs. \cite{ehrnsperger95b} and 
\cite{gehrmann95a}. These parton distributions
have been slightly modified because we use 
different unpolarized parton distributions
in both cases. We have taken those from 
Gl\"uck et al. \cite{glueck90a}.
If the helicity distributions are required
to be positive, the absolute value of the 
polarized distributions cannot be larger 
than the unpolarized ones, viz.
\begin{equation}
\label{positivity}
f_\pm(x)=\frac{1}{2}(f(x)\pm\Delta f(x))>0 
\; \Rightarrow
|\Delta f(x)|\leq f(x) \;\;\;\; (f=q,g).
\end{equation}
Whenever $|\Delta f(x)|$ gets larger than
$f(x)$ we substitute it by  
$|\Delta f(x)|=f(x)$. 
This modifies the parton distributions 
a little bit at large x as can be seen from 
fig. \ref{geh}.
\\
The first parametrization of the polarized
parton distributions \cite{ehrnsperger95b} 
is based on a Carlitz-Kaur type 
spin-dilution model for the proton's 
distribution functions. It is referred to as 
CKT. The important aspects of this model are:
\begin{itemize}
\item The gluon distribution can be expressed
      through the valence quarks because of
      the spin-dilution idea.
\item It uses a non-standard ratio of the 
      SU(3) coupling constants 
      $F/D=0.49\pm0.08$. 
\item The free parameters of the model
      are chosen to fulfill the Bj{\o}rken 
      and Ellis-Jaffe sum rule which turns
      out to be possible due to the above 
      non-standard $F/D$ ratio.
\item The fit to the $g_1(x,Q^2)$ data is made 
      at $Q_0^2=10\,{\rm GeV}^2$. There are
      no polarized strange quarks at this $Q^2$. 
\item The first moment of the gluon distribution
      $\Delta g=0.262$ is rather small in 
      this model. So it makes no difference if
      one uses the anomaly relation 
      (\ref{anomalycontrib}) or the naive 
      identification.
\end{itemize}
The shape of these parton distributions can best 
be read off from fig. \ref{bruno}. They are shown
in this figure for a low resolution of $Q=2$\,GeV 
as well as for a large resolution that is 
approximately equal to the $W^\pm$ mass. 
Note that we use $n_f=3$ for the number of
flavors in the GLAP equation, whereas $n_f=2$ has 
been taken in ref. \cite{ehrnsperger95b}.    
\\
The second type of parton distributions,
taken from \cite{gehrmann95a}, is a simple fit to 
$g_1$ using $\beta$-decay and hyperon decay
data to fix the first moments of the polarized 
distributions. 
For the shape of the distributions 
a generic function is assumed and the shape 
parameters are fitted to the $g_1$ data.
The important aspects of this distributions are:
\begin{itemize}
\item The sea is taken to be SU(3) flavor
      symmetric and vanishes at 
      $Q_0^2=4\,{\rm GeV}^2$.
\item The gluon polarization 
      $\Delta g\,= 1.971$ is rather large, 
      mainly due to the larger value
      $F/D=0.590$ and the anomaly 
      equation (\ref{anomalycontrib}).
\item Three different shapes of the distributions
      are considered. They all fit the 
      data equally well. These parton 
      distributions are called SETA, SETB, and
      SETC.
\end{itemize}
The shapes of the the distributions can be read off from 
fig. \ref{geh}. It can also be seen from this
figure that the valence quarks in both parton 
distributions are quite similar in size and shape.
\section{Results}
\begin{figure}
\centering{\psfig{figure=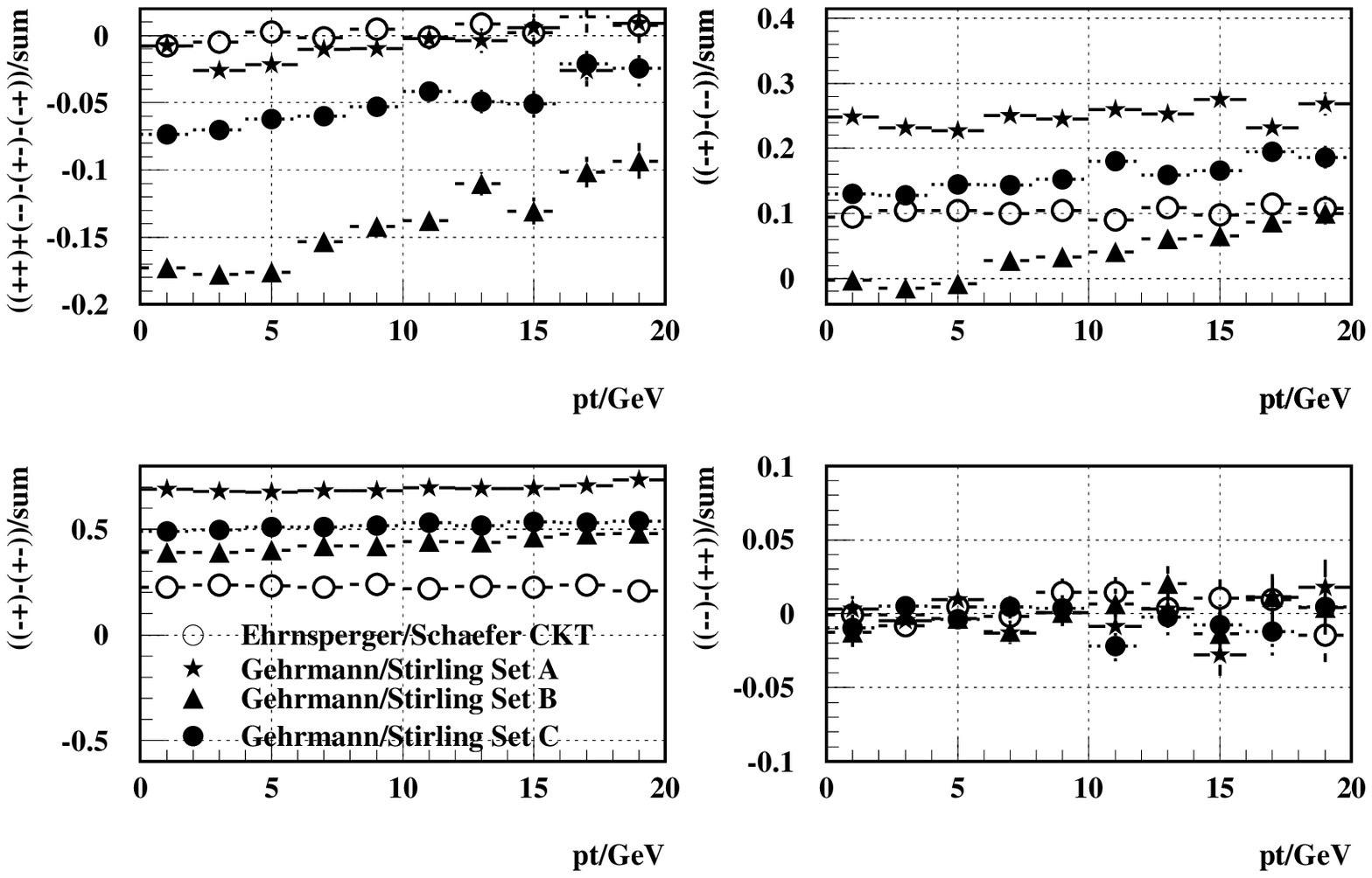,height=20cm}}
\caption{Transverse momentum distributions of the 
spin asymmetries defined in eqns. 
(\protect\ref{allpc}), (\protect\ref{allpv1}),
(\protect\ref{allpv2}), and 
(\protect\ref{allpv3}) for  
$W^+$-production $2\rightarrow 2$ 
processes and  SETA, SETB, SETC, and CKT partons.}
\label{w+22pt}   
\end{figure}

\begin{figure}
\centering{\psfig{figure=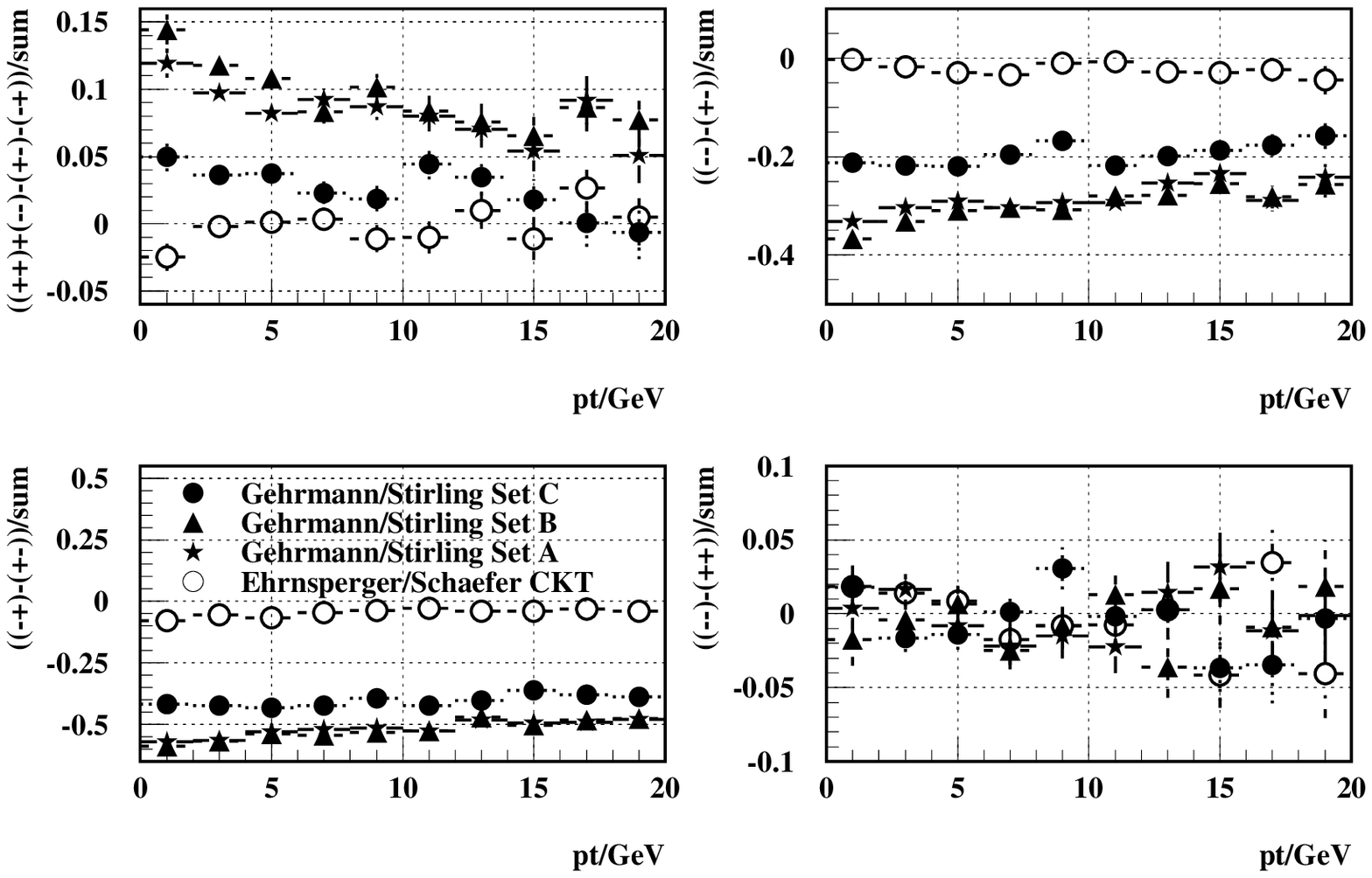,height=20cm}}
\caption{Transverse momentum distributions of the 
spin asymmetries defined in eqns. 
(\protect\ref{allpc}), (\protect\ref{allpv1}),
 (\protect\ref{allpv2}), and (\protect\ref{allpv3}) for 
the $W^-$-production $2\rightarrow 2$ processes and 
SETA, SETB, SETC, and CKT partons.}   
\label{w-22pt}
\end{figure}

\begin{figure}
\centering{\psfig{figure=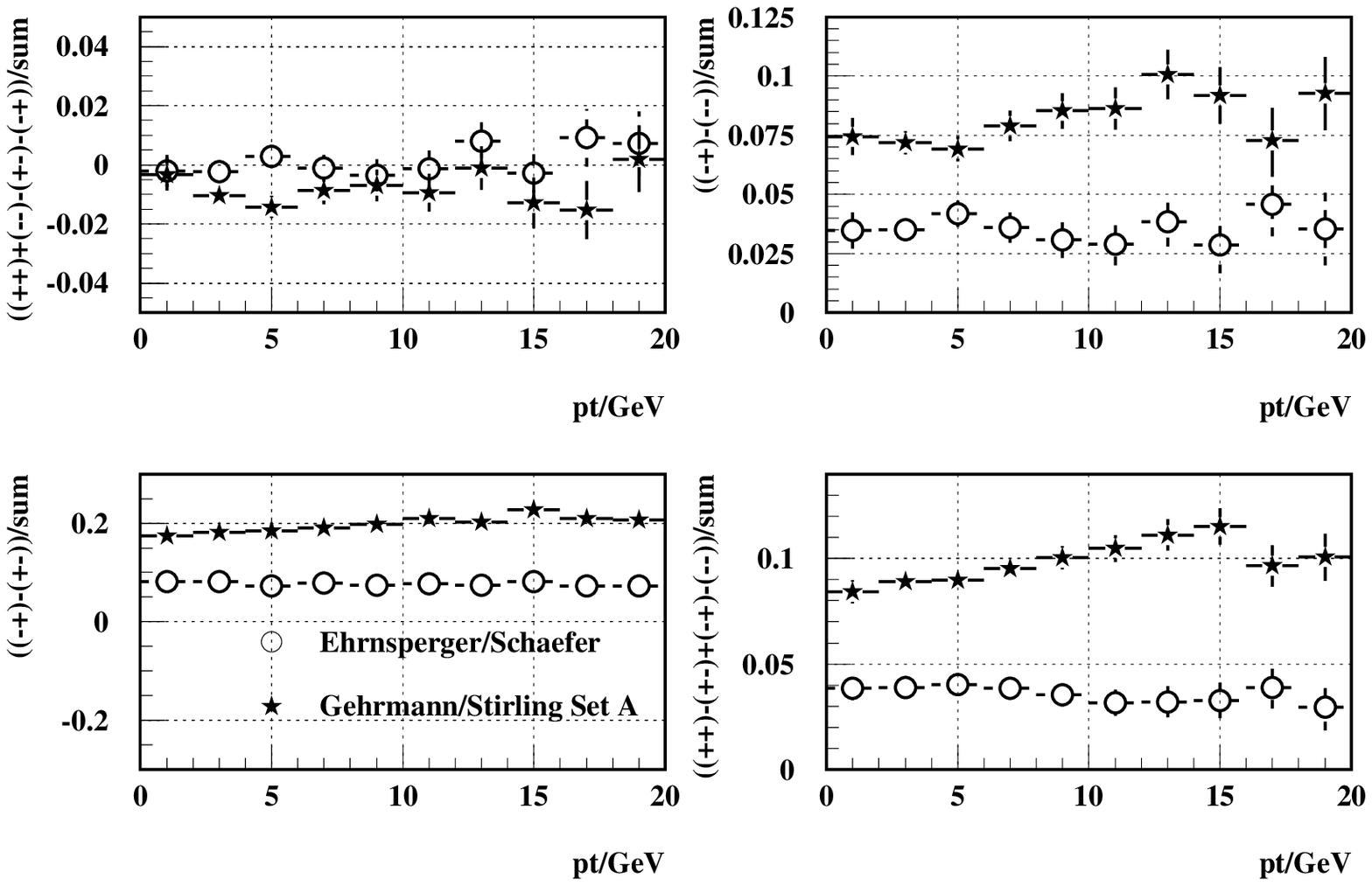,height=20cm}}
\caption{Transverse momentum distributions 
of the spin asymmetries defined in eqns. 
(\protect\ref{allpc}), (\protect\ref{allpv1}),
\,(\protect\ref{allpv2}), and (\protect\ref{allpv3}) 
for the $Z_0$-production $2\rightarrow 2$ processes 
and SETA and CKT partons.}
\label{z22pt}  
\end{figure}

\begin{figure}
\centering{\psfig{figure=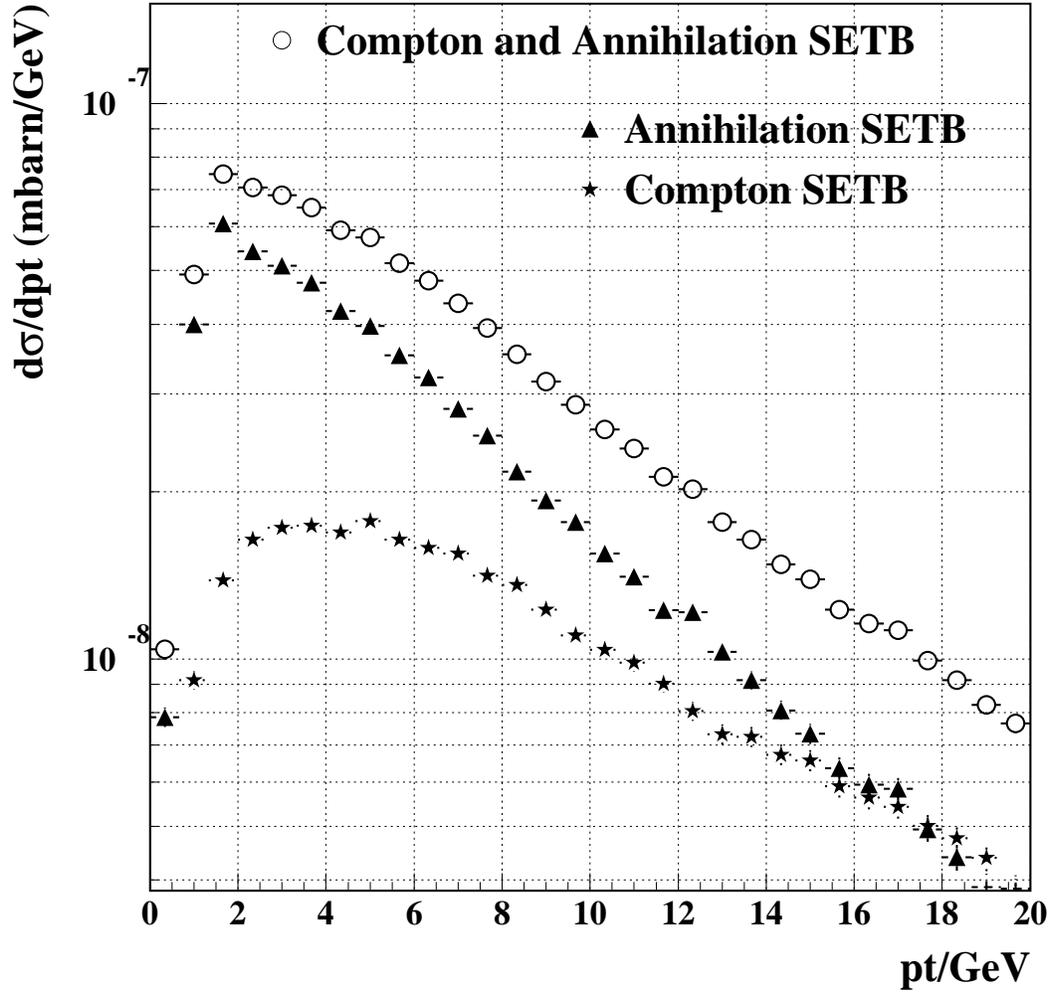,height=20cm}}
\caption{Contributions to the transverse momentum 
distribution of $W^+$'s for a spin orientation
$(-\,+)$ of the protons and SETB partons. This
spin orientation gives the largest ratio of Compton
to annihilation scattering.}
\label{w+22ca6-+}  
\end{figure}

\begin{figure}
\centering{\psfig{figure=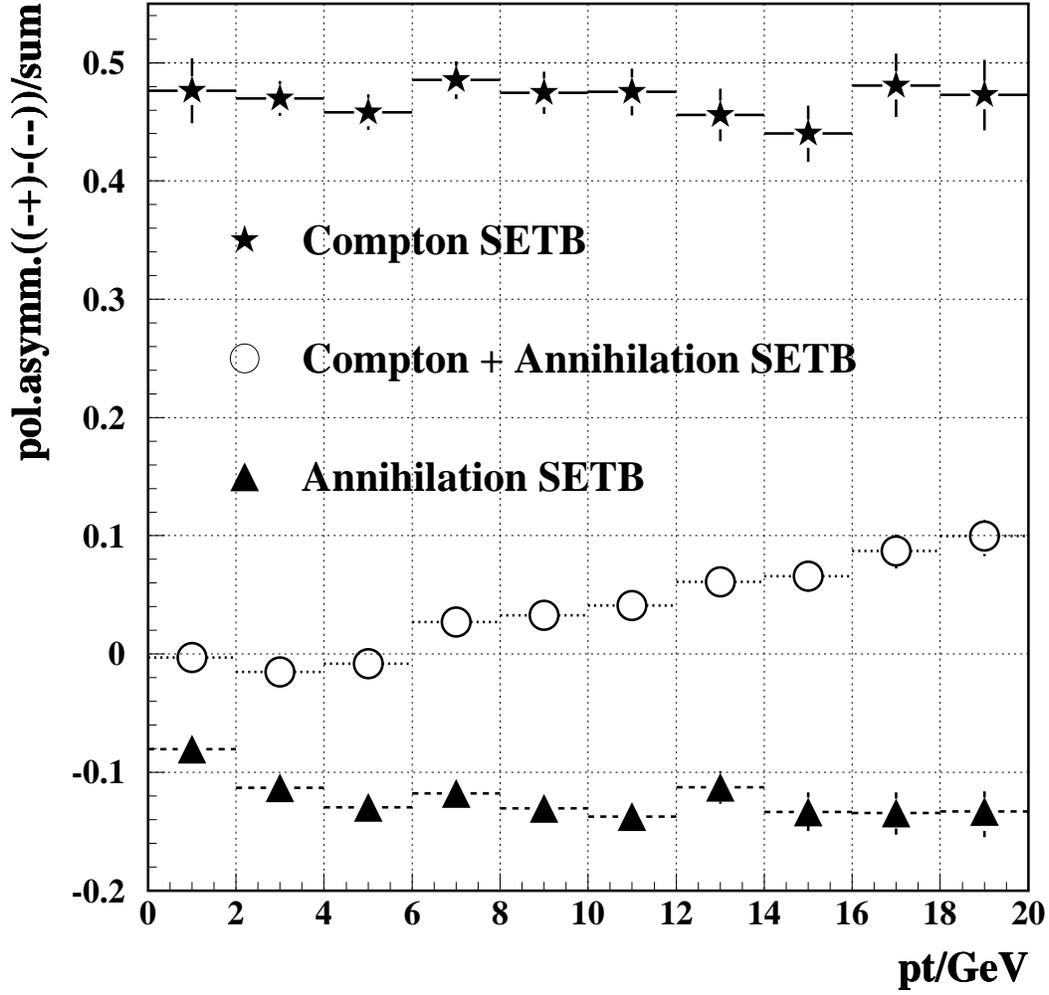,height=20cm}}
\caption{Contributions to the asymmetry  
$A_{LL}^{PV2}$ in $W^+$ production for SETB partons.}
\label{w22+6c}  
\end{figure}

\begin{figure}
\centering{\psfig{figure=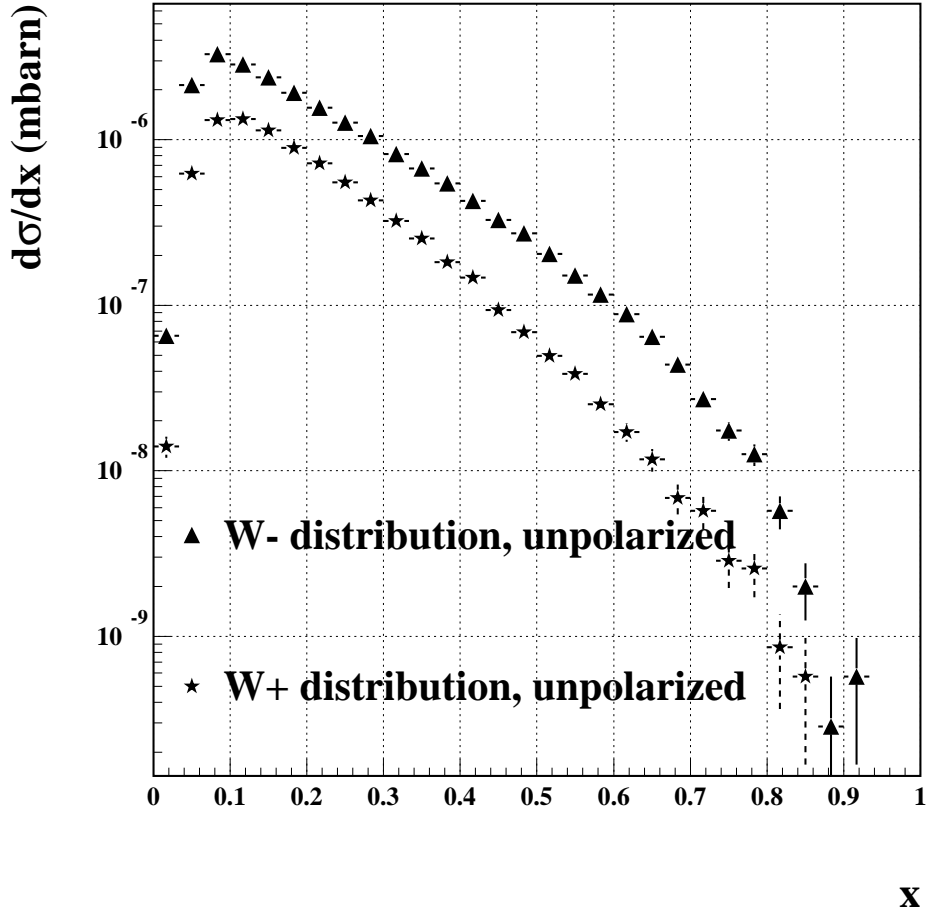,height=20cm}}
\caption{The $x$ distribution of  
$W^\pm$'s for unpolarized proton scattering at
$\protect\sqrt{s}=500$\,GeV. Only events with 
\protect$p_T>5$\,GeV have been considered.}
\label{xhptmain}  
\end{figure}

Before showing our results, we have to fix
some notation and introduce the spin 
asymmetries. By $(s_1,s_2),\;\;s_1,s_2=\pm$, 
we will denote a {\em spin} configuration of the 
scattering protons. $s_1=+$ means that the proton
which moves in the direction of the positive
z-axis has a spin that also points in the 
z-direction. $s_1=-1$ means that the spin 
of this proton points in the negative z-direction.
The spin of this particle is by 
convention equal to its helicity.  
The spin of the other proton 
is denoted by $s_2$. Note that its
helicity is always opposite to its spin, e.g.,
$(+,+)$ means that the `beam proton' has spin and
helicity $+$, whereas the `target proton' has
spin $+$ and helicity $-$.
\\
By a $2\rightarrow 1$ process we mean resonance 
production through a hard scattering cross 
sections (\ref{wresonance}), and 
(\ref{zresonance}). A $2\rightarrow 2$ process 
proceeds by definition through the annihilation 
amplitude (\ref{annihilation}), or Compton amplitude
(\ref{compton}). In both cases we generate
higher order corrections through the MC 
algorithm explained in section \ref{sphinx}.
This means in particular that 
$2\rightarrow 1$ processes together with  radiation 
contain those parts of $2\rightarrow 2$ processes 
that survive the LLA. Therefore, one should switch on 
either $2\rightarrow 1$ processes or  
$2\rightarrow 2$ process, not both. For large
transverse momenta $2\rightarrow 2$ processes 
must be used because they describe
the emission of an additional parton more 
accurately. On the other hand, if the transverse
momentum is low, not any event is accompanied
by an additional parton radiation and one should
switch to the $2\rightarrow 1$ processes.
If we compare the cross sections, we find
that the switch from  $2\rightarrow 1$ to  
$2\rightarrow 2$ processes should be made for 
transverse momenta of about $p_T=4\,{\rm GeV}$. 
The same $p_T$  is obtained by comparing the 
asymmetries derived from $2\rightarrow 1$ and
$2\rightarrow 2$ processes. 
\\
After these preliminary remarks we define the 
asymmetries and show the results. The parity 
conserving asymmetry is defined through
\begin{equation}
\label{allpc}
A_{LL}^{PC}=\frac{(+\,-)+(-\,+)-(+\,+)-(-\,-)}
{(+\,-)+(-\,+)+(+\,+)+(-\,-)}.   
\end{equation}
We consider three different parity violating
asymmetries, namely, 
\begin{equation}
\label{allpv1}
A_{LL}^{PV1}=\frac{(+\,-)-(-\,+)}{(+\,-)+(-\,+)},
\end{equation}
\begin{equation}
\label{allpv2}
A_{LL}^{PV2}=\frac{(-\,+)-(-\,-)}{(-\,+)+(-\,-)},
\end{equation}
\begin{equation}
\label{allpv3}
A_{LL}^{PV3}=\frac{(-\,-)-(+\,-)}{(-\,-)+(+\,-)}.
\end{equation}
A test asymmetry 
\begin{equation}
\label{al}
A_{LL}^{TEST}=\frac{(++)-(-\,-)}
{(++)+(-\,-)},
\end{equation}
which vanishes trivially from rotation invariance,
shows the quality of the calculation.
\\
The spin-asymmetries are plotted in figs. 
\ref{w+22pt}, \ref{w-22pt}, and \ref{z22pt} 
against the transverse momentum of the vector 
bosons in the CMS frame of the
reaction. For each orientation of the protons, 
0.1 million events have been produced 
\footnote {This requires about 3 hours of CPU time 
for one spin orientation on an IBM-RISC6000 workstation.}. 
This is a roughly realistic number for the RHIC
spin experiment with a luminosity of $10^{32}/cm^2 s$ 
and $1.5$ years of running.
Some asymmetries are quite sensitive 
to the structure functions. But one must be 
careful not to attribute this to the 
{\em size} of the polarized gluon distribution 
alone. First of all, one can see from the 
asymmetries corresponding to SETA, SETB, and SETC, 
which only differ by the shape of the gluon 
distribution, that the magnitude of the 
asymmetries is quite sensitive to the {\em shape} 
of $\Delta g$. See fig. \ref{w+22pt} and note that 
all Gehrmann/Stirling distributions (full symbols)
have the same $\Delta g$ but give very different
asymmetries. These asymmetries overlap with those 
obtained for the CKT distributions (open circles), 
which have much smaller $\Delta g$. 
Interestingly, there is another significant 
distinction between the distributions with small 
and large $\Delta g$:
Some asymmetries derived from SETA, SETB, and SETC
vary significantly with transverse 
momentum $p_T$, whereas those derived from the
CKT distributions stay approximately constant. 
One can see this most clearly from the parity 
violating asymmetries $A_{LL}^{PV2,W^+}$ and 
$A_{LL}^{PV3,W^-}$. This effect can be traced back
to the hard scattering cross sections eqns. 
(\ref{annihilation}), and (\ref{compton}). Even 
though the Compton process is suppressed by
a factor of $3/8$, it can become comparable 
to the annihilation process for large transverse
momenta because the Compton scattering falls
off less rapidly than the annihilation scattering.
At very large transverse momenta the Compton
induced asymmetry will saturate the spin
asymmetry. Fig. \ref{w+22ca6-+} shows that
the transition region from an annihilation 
induced asymmetry to a Compton induced asymmetry
can be reached for the distribution SETB (and 
as well for SETA, and SETC). This is not possible
for the CKT distribution, at least up to 
$p_T\approx 20$\,GeV. The transition can 
also be observed in the corresponding 
spin asymmetry fig. \ref{w22+6c}.
Finally, one can see from fig. \ref{xhptmain} 
that the range in x where the cross section is
large lies approximately between 
$0.1\stackrel{<}{\sim}x\stackrel{<}{\sim}0.5$
with a strong maximum at about $x=0.1$. 
\section{conclusions}
We have presented an extension of the MC generator
SPHINX to polarized $W^\pm$ and $Z_0$ production.
Using two different types of parton distributions
in a MC simulation at ${\sqrt s}=500$\,GeV, 
we find that several details of the calculated 
asymmetries show an unexpected sensitivity to the
polarized parton distributions. The variation of
the spin asymmetries  $A_{LL}^{PV2,W^+}$ and 
$A_{LL}^{PV3,W^-}$ with transverse momentum
$p_T < 20$\,GeV are especially interesting.
The upper limit in $p_T$ in our simulation
is dictated by MC statistics. 
For $p_T > 4$\,GeV we use NLO hard scattering 
cross sections in combination with initial state 
radiation. The simulations show that not only the 
size but also the $p_T$ dependence of some 
asymmetries is sensitive to $\Delta g$.
To make use of this information, one has to
measure the total $p_T$ of the hadronic products
in addition to the charged lepton from the $W^\pm$ 
decay.
\\
We only showed results for the transverse momentum
spectra because they are especially interesting for
measuring the polarized gluon distribution. There 
many other important observables that can be 
generated by SPHINX. Rapidity distributions, for 
example, are sensitive to flavor asymmetries in the 
polarized sea quark distribution. 
\acknowledgements
We thank B. Ehrnsperger, S. G\"{u}llenstern,
L. Mankiewicz, and E.Stein  for illuminating 
discussions, and S. Gehrmann for providing us 
with his parton distributions. A. Sch\"{a}fer
acknowledges support by DFG (G. Hess Programm),
BMBF, and MPI f\"{u}r Kernphysik (Heidelberg).
\newpage
\newpage
\newpage
\newpage
\newpage
\newpage
\end{document}